\documentclass[prb,twocolumn,showpacs,preprintnumbers]{revtex4}

\def\id{{\rm d}}

\def\eh{\left(\frac{2e}{\hbar}\right)}
\def\he{\left(\frac{\hbar}{2e}\right)}

\def\be{\begin{equation}}
\def\ee{\end{equation}}
\def\bea{\begin{eqnarray}}
\def\eea{\end{eqnarray}}
\def\nn{\nonumber\\}

\usepackage{graphicx}
\usepackage{dcolumn}
\usepackage{bm}

\begin{document}


\title{A theoretical analysis of flux-qubit measurements with a dc-SQUID}

\author{Hayato Nakano$^{1,3}$, Hirotaka Tanaka$^{1,3}$, Shiro Saito$^{1,3}$, 
Kouichi Semba$^{1,3}$, Hideaki Takayanagi$^{1,3}$, and Masahito Ueda$^{2,3}$}
\affiliation{
$^{1}$NTT Basic Research Laboratories, NTT Corporation, Atsugi-shi, Kanagawa 243-0198, Japan.\\
$^{2}$Tokyo Institute of Technology, Meguro-ku, Tokyo 152-8551, Japan.\\
$^{3}$CREST, Japanese Science and Technology Agency}%

\begin{abstract}
The readout process of a superconducting flux-qubit is theoretically analyzed
in terms of the quantum dynamics of a qubit-SQUID coupled system during
measurement. The quantity directly observed by the measurement is the switching current 
($I_{\rm sw}$) of the dc-SQUID placed around the qubit ring.
In order to clarify the relation between the $I_{\rm sw}$ and the qubit state,
we calculated the time evolution of the density operator of the qubit-SQUID system
while increasing the SQUID bias current until switching events occur and obtained the switching current
distributions. This clarifies what information of the qubit is obtained
by a dc-SQUID switching current measurement
under specific conditions, for example,
when the qubit eigenstate is a superposition of two different flux states. 
\end{abstract}

\maketitle

\section{Introduction}

The quantum bit (qubit) is the fundamental element of quantum computers\cite{N-C}.
Recently, many experiments have been reported on quantum two-state systems consisting
of superconducting circuits with Josephson junctions\cite{Friedman,Casper,Mar,WalT,Irinel,Naka,Yu,Vion,Tanaka}.
Some of these systems use  two flux-states in the superconducting ring that are macroscopically
distinguishable from each other, and the superposition of these two states
\cite{Friedman,Casper,WalT,Irinel,Tanaka}. These are called
superconducting flux-qubits. The two states can be
characterized by the directions of the circulating supercurrent along the ring, and the magnetic
flux that is induced by the supercurrent.

In experiments, small magnetic-field measurements
made with dc-SQUIDs are used to read out  the flux qubit states.
A dc-SQUID is a highly sensitive magnetic flux probe.
The measurement result is the bias current at which the
SQUID switches to the voltage state.
The switching current ($I_{\rm sw}$) of the SQUID varies
depending on the quantum state of the qubit. 

The qubit ring has two stable states. One induces the flux $+\phi_{\rm q}$, and the other $-\phi_{\rm q}$.
The SQUID switches at a bias current of $I_{\rm sw\pm}=I_{\rm c0}\cos[(\Phi_{\rm SQ}\pm \phi_{\rm q})/\Phi_0]$
depending on which state the ring is in. Here, $\Phi_{\rm SQ}$ is the external flux applied to the SQUID ring,
and $\Phi_{0}=h/(2e)$ is the flux quantum. Therefore,
at $\Phi_{\rm SQ}/\Phi_{0}=1/2$, where the two states are energetically degenerated,
the switching current appears probabilistically  at $I_{\rm sw+}$ {\it or} $I_{\rm sw-}$ 
for each measurement
in the sense of classical statistics.

When Josephson junctions in the ring
are small enough the ring becomes a quantum two-state system, that is, a qubit.
Then, it is not so trivial
what is meant by the obtained switching current, especially 
when the qubit is not in an eigenstate of 
the quantity measured by the SQUID.

There are at least two possibilities. A first possibility is that
the dc-SQUID measurements of flux-qubits is a projection measurement of the small flux
induced by the qubit ring current. The qubit has only two states where the current is definite.
The switching current appears probabilistically at $I_{\rm sw+}$ {\it or} $I_{\rm sw-}$
even when the qubit state is a superposition of the two possible states before the measurement.
Then, after the measurement, the qubit state jumps to the one of the two states corresponding to the measurement
result. A second possibility is that the switching current has an intermediate value between 
$I_{\rm sw+}$ and $I_{\rm sw-}$ and does not behave probabilistically
even for a qubit that is in a superposition state.

In experiments, when the fluctuation of the switching current is larger than $|I_{\rm sw+}-I_{\rm sw-}|$
and when we need to take the average of many measured switching current values in order to 
determine the qubit-inducing flux, it is impossible to distinguish 
which  of
the two possibilities above is correct because the averaged values always  appear between  $I_{\rm sw+}$ and $I_{\rm sw-}$
independent of the possibilities and never behave probabilistically. 
To clarify what each switching current obtained by each measurement corresponds to,
and what is meant by the distribution of a huge number of measured switching currents,
we must undertake a quantitative analysis of the dynamics of the whole system,
which consists of the measured object (flux qubit) and the measurement apparatus (dc-SQUID).


The rest of this paper is organized as follows.
Section II briefly introduces the background to superconducting flux-qubit measurement
with a dc-SQUID. Section III shows the Hamiltonian of the flux-qubit and dc-SQUID coupled system, and
derives a simplified version, which is applicable to numerical calculations.
Section IV describes  our numerical calculation of the 
time evolution of the  density operator of the qubit-SQUID composite system.
Section V shows the results of the calculations and discusses them. Section VI makes some concluding remarks.

\section{Flux-qubit and reading out of  states with a dc-SQUID.}

\subsection{Superconducting flux-qubits}

\begin{figure}%
\includegraphics[width=5.5cm]{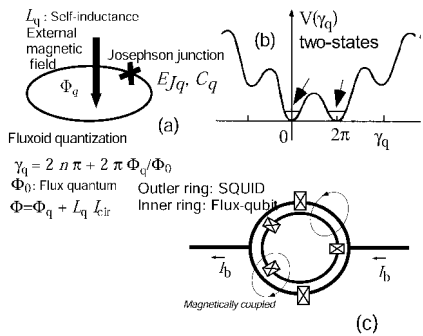}
\caption{\small 
Principle of the flux qubit.
(a) Superconducting ring and fluxoid quantization.
(b) Potential energy for the Josephson phase  $\gamma_{\rm q}$ and the two states.
(c) Geometry of the dc-SQUID and the flux qubit for the measurement setup.
}
\label{flux-qubit}
\end{figure}

A superconducting flux qubit is a superconducting ring interrupted by
a Josephson junction (or junctions).
The fluxoid quantization causes the junction phase difference $\gamma_{\rm q}$ and the external
flux $\Phi_{\rm q}$ applied by the magnetic field
to obey the condition such that 
\be
\gamma_{\rm q}=2 n \pi + 2 \pi \frac{\Phi}{\Phi_0},
\ee
where $\Phi=\Phi_{\rm q}+L_{\rm q} I_{\rm cir}$, $n$ in an integer,
and $I_{\rm cir}$ is the circulating supercurrent along the ring
[Fig. \ref{flux-qubit}(a)].

Suppose that an external flux $\Phi_{\rm q} \sim (n+1/2) \Phi_0$ pierces the qubit ring.
The two states in the wells, $\Phi_{\rm q} \sim n \Phi_0$ and $\Phi_{\rm q} \sim (n+1) \Phi_0$,
of the potential energy $V(\gamma_{\rm q})$ are energetically degenerate
[Fig. \ref{flux-qubit}(b)]. We denote these two states  by $|{\rm L}\rangle$ and $|{\rm R}\rangle$.
The
Josephson plasma oscillation energy $\hbar \omega_{\rm p} \sim \sqrt{2 E_{\rm C}E_{\rm J}}$
gives approximate level splitting in each well, where $E_{\rm C}\equiv (2e)^2/(2 C_{\rm q})$ and $E_{\rm J}$ are the charging energy and the Josephson energy of
the junction, respectively.
When the potential barrier at $\Phi_{\rm q} \sim (n+1/2) \Phi_0$ is comparable
to the charging energy $E_{\rm C}$, 
macroscopic quantum tunneling of the phase couples states between
the barrier.  When the anti-crossing energy separation between the coupled states
are much smaller than the level splitting $\hbar \omega_{\rm p}$,
two states, those are energetically nearest and in different wells, are almost
independent other states. Then, we can consider these two states
as a two-state system, that is, a qubit.


We can treat the states of the ring as a pseudo two-level system and obtain 
a reduced two-level-system Hamiltonian with a 
$\{\left|{\rm L} \right\rangle ,\left|{\rm R} \right\rangle \}$ basis: 
\begin{equation}
\label{eq1}
H_{\rm q} = \varepsilon \sigma _z - \Delta \sigma _x ,
\label{Hq01}
\end{equation}
\noindent
where $2 \varepsilon $ is the energy difference between the localized states 
in the left and right wells and can be controlled by $f$. Here, $\Delta$ 
represents the transfer energy due to macroscopic quantum tunneling between the wells, and 
\bea
\label{eq2}
\displaystyle{
\sigma _z = \left|{\rm L} \right\rangle \left\langle L \right| - \left|{\rm R} 
\right\rangle \left\langle R \right| \equiv \left( {{\begin{array}{*{20}c}
 1 \hfill & 0 \hfill \\
 0 \hfill & { - 1} \hfill \\
\end{array} }} \right), }\nn
\displaystyle{\sigma _x = \left|{\rm L} \right\rangle \left\langle R \right| + \left|{\rm R} 
\right\rangle \left\langle L \right| \equiv \left( {{\begin{array}{*{20}c}
 0 \hfill & 1 \hfill \\
 1 \hfill & 0 \hfill \\
\end{array} }} \right).}
\eea

The eigenenergy of the ground (excited) states of Hamiltonian \ref{Hq01} is given by 
$E_{\rm g(e)} = - ( + )\sqrt {\varepsilon ^2 + \Delta ^2} $. The eigenstates are 
\begin{eqnarray}
\label{eq3}
\left| {\rm g} \right\rangle = \sin \left[ \frac{\theta}{2} \right]\left|{\rm L} 
\right\rangle + \cos \left[ \frac{\theta}{2} \right]\left|{\rm R} \right\rangle , \nn
\left|{\rm e} \right\rangle = - \cos \left[ \frac{\theta}{2} \right]\left|{\rm L} 
\right\rangle + \sin \left[ \frac{\theta}{2} \right]\left|{\rm R} \right\rangle, 
\end{eqnarray}
where $\tan[\theta]=\Delta/\varepsilon$.
The energy levels exhibit anti-crossing at $\Phi_{\rm q}=(n+1/2)\Phi_0$ due 
to the mixing of $|{\rm L}\rangle$ and $|{\rm R}\rangle$ via
macroscopic quantum tunneling.
The quantum mechanical {\rm average} of the qubit circulating current 
$\langle I_{\rm cir} \rangle= I_{\rm co} \left\langle i \right|\sigma _z \left| i \right\rangle $ 
$(i = {\rm g, e})$, for the ground $\left| {\rm g} \right\rangle $ and the first excited 
states $\left|{\rm e}\right\rangle $, can be obtained using $\left\langle {\rm g} 
\right|\sigma _z \left| {\rm g} \right\rangle = - \cos \theta$ and 
$\left\langle{\rm e} \right|\sigma _z \left| {\rm e} \right\rangle = \cos \theta $,
where $I_{\rm c0} $ is the maximum supercurrent along the 
qubit ring.

\subsection{Measurement with dc-SQUID}

The ``readout" of the flux qubit is the result of
measurement to determine whether the qubit is in the $|{\rm L}\rangle$ or $|{\rm R}\rangle$ state.
The difference between them,
which is measurable from the outside, is the direction of the circulating supercurrent along the ring
and the magnetic field induced by the current.
The field changes its sign depending on the states $|{\rm L}\rangle$ and $|{\rm R}\rangle$.
Then, we expect that, by measuring  the field with a dc-SQUID, we can distinguish  the state of the qubit.
This method is actually used in the experiments.

The switching current $I_{\rm sw}$ of the SQUID is the bias current value
at which the sum  of the macroscopic phase differences of two junctions,
$\gamma_{0+}/2 \equiv (\gamma_1+\gamma_2)/2$,
escapes from the potential well.
In a classical situation, where the external flux $\Phi_{\rm SQ}$ is added and a small flux $\pm \phi_{\rm q}$
is induced in the SQUID ring,
the switching current is
$I_{\rm sw}(\pm \phi_{\rm q})\sim I_0(\cos[\pi \Phi_{\rm SQ}/\Phi_0]\mp \frac{\pi \phi_{\rm q}}{\Phi_0}\sin[\pi \Phi_{\rm SQ}/\Phi_0])$,
where $I_0\equiv \eh E_{\rm J0}$, and $E_{\rm J0}$ is the Josephson coupling energy of the junction.
In real qubit measurements, the induced flux is  $10^{-3}\Phi_0 \sim 10^{-2}\Phi_0$.

This type of quantum superposition state is called as macroscopic quantum 
coherence (MQC). The question of whether MQC is actually possible, was 
originally introduced by A. J. Leggett, in the 1980's \cite{C-L}. Recently MQC has been experimentally 
observed in a 
superconducting ring with Josephson junctions\cite{Irinel}. 
This MQC can also be used as a qubit because it is a typical 
two-state quantum system with controllable parameters. 
A qubit based on MQC in a superconducting ring is called a 
superconducting flux-qubit. 

In experiments,
the dc-SQUID is fabricated out of the qubit ring
in order to measure a small magnetic field induced
by the qubit circulating current [Fig. \ref{flux-qubit}(c)].
The external field is tuned so that the flux piercing the qubit
was set near the degenerate point $\Phi_{\rm q}=(n+1/2)\Phi_0$.

\section{Derivation of simplified Hamiltonian for the calculations}

\subsection{Total Hamiltonian for the qubit and dc-SQUID coupled system}

\begin{figure}
\begin{center}
\includegraphics[width=5.5cm]{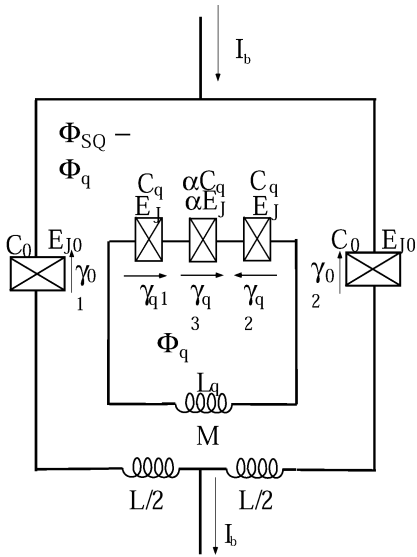}
\end{center}
\caption{
Quantum circuit for the flux-qubit and dc-SQUID coupled system.}
\label{NTT}
\end{figure}

The real flux-qubit used in our experiments has three Josephson junctions. 
The dc-SQUID with two Josephson junctions has two leads. The ring and the dc-SQUID are coupled magnetically 
via mutual inductance $M$. This structure was  proposed by  
Mooij's group at the Technical University of Delft \cite{Mooij}.

The superconducting circuit for the total system that consists of the flux-qubit ring and the dc-SQUID 
is shown in Fig. \ref{NTT}. The Hamiltonian is given by
\begin{equation}
H=H_{\rm SQ}+H_{\rm q}+H_{\rm int},
\label{Tot}
\end{equation}
where
\begin{eqnarray}
H_{\rm q}=\frac{C_{\rm q}}{4}\he^2\left({\dot{\gamma}_{\rm q+}}^2+{\dot{\gamma}_{\rm q-}}^2
+2 \alpha{\dot{\gamma}_{\rm q3}}^2 \right) \nn
-2 E_{\rm J}\cos[\gamma_{\rm q+}/2]\cos[\gamma_{\rm q-}/2] 
-\alpha E_{\rm J}\cos[\gamma_{\rm q3}] \nn
+\he^2\frac{L(\gamma_{\rm q-}
+\gamma_{\rm q3}-2 \pi f_{\rm q}
)^2}{2(L_{\rm q}L-M^2)},
\end{eqnarray}
is the qubit Hamiltonian,
\begin{eqnarray}
H_{\rm SQ}=
\frac{C_{\rm 0}}{4}\he^2\left({\dot{\gamma}_{\rm 0+}}^2+{\dot{\gamma}_{\rm 0-}}^2\right) \nn
-2 E_{\rm J0}\cos[\gamma_{\rm 0+}/2]\cos[\gamma_{\rm 0-}/2] -\he I_{\rm b}(t)\frac{\gamma_{+}}{2} \nn
+\he^2\frac{L_{\rm q}\left(\gamma_{\rm 0-}-2\pi f_{\rm SQ}\right)^2}{2(L_{\rm q}L-M^2)},
\end{eqnarray}
is the SQUID Hamiltonian, where $I_{\rm b}(t)$ is the bias current, which increases in time from zero to
above the switching slowly enough compared with other relevant time scales, and
\be
H_{\rm int}=\he^2\frac{M\left(\gamma_{\rm 0-}-2\pi f_{\rm SQ}\right)
\left(\gamma_{\rm q-}+\gamma_{\rm q3}-2\pi f_{\rm q}\right)}{(L_{\rm q}L-M^2)},
\ee
is the interaction between the qubit and the SQUID.
Here, notations such as,
\begin{equation}
\gamma_{\rm q\pm}\equiv \gamma_{\rm q1}\pm\gamma_{\rm q2},
\hspace{1mm}
\gamma_{\rm 0\pm}\equiv \gamma_{\rm 01}\pm\gamma_{\rm 02},
\end{equation}
\begin{equation}
f_{\rm q}\equiv \Phi_{\rm q}/\Phi_0, 
\hspace{1mm} 
f_{\rm SQ}\equiv \Phi_{\rm SQ}/\Phi_0,
\end{equation}
are adopted.
Moreover, 
$L$ and $L_{\rm q}$ are the self-inductance of the SQUID ring and the qubit ring, respectively.
$M$ is the mutual inductance between them. $C_{\rm q}$ and  $E_{\rm J}$ are the capacitance and the
Josephson energy of two of three junctions in the qubit, and the other junction has values of 
$\alpha C_{\rm q}$ and  
$\alpha E_{\rm J}$ with $\alpha\sim 0.8$. $C_{\rm 0}$ and  $E_{\rm J0}$ are the capacitance and the
Josephson energy of the two junctions in the SQUID.



\subsection{Effective interaction between $\gamma_{0+}$, and $\sigma$}

\subsubsection{Elimination of variables, $\gamma_{\rm 0-}$ and $\gamma_{\rm q3}$}

The Hamiltonian Eq. (\ref{Tot}) of the total system contains five quantum variables;
$\gamma_{0+}$, $\gamma_{0-}$, $\gamma_{\rm q+}$, $\gamma_{\rm q-}$, and $\gamma_{\rm q3}$.
However,
the quantity we really measure is the switching current $I_{\rm sw}$,
that is, the SQUID bias current at which
$\gamma_{0+}$ escapes from the Josephson potential well.
Therefore, we have to analyze the way in which  the relationship between
the $\gamma_{0+}$ of the SQUID and the qubit state evolves in time 
in the presence of  the SQUID-qubit interaction.

The qubit state is mainly determined by the value of $\gamma_{\rm q-}$, as described
below. 
In terms of the qubit measurement, important variables are $\gamma_{\rm q-}$ and
$\gamma_{\rm 0+}$. The other variables are not directly accessible 
quantities that can neither be controlled nor measured
directly. Therefore, it is desirable that these  variables are eliminated thus allowing
the total system to be analyzed as a two-variable system.

Fortunately, $\gamma_{0-}$ and  $\gamma_{\rm q3}$ can be approximated as harmonic
oscillators strongly confined in parabolic potentials. We can integrate out 
these variables with the path-integral method, as described in Appendix A. 
Moreover, since the energy-level spacing
of these harmonic oscillators is very wide, the contributions, which are non-local in time,
are negligible and we obtain a simplified Hamiltonian for the total system
as 
\be
H'=H'_{\rm q}+H'_{\rm SQ}+H'_{\rm int}
\ee
where
\bea
H'_{\rm q}&=&\he^2\frac{C_{\rm q}}{4}\left(\dot{\gamma}_{\rm q+}^2+\dot{\gamma}_{\rm q-}^2
+2 \alpha \dot{\gamma}_{\rm q-}^2\right)\nn
&&-2 E_{\rm J}\cos[\gamma_{\rm q+}/2]\cos[\gamma_{\rm q-}/2]
-\alpha E_{\rm J}\cos[\gamma_{\rm q-}-2 \pi f_{\rm q}],\nn
\label{SHq}
\eea
\bea
H'_{\rm SQ}=\he^2\frac{C_0}{4}\dot{\gamma}_{\rm 0+}^2
&-&2 E_{\rm J0}\cos[\gamma_{\rm 0+}/2]\cos[\pi f_{\rm SQ}]\nn
&-&\he I_{\rm b}(t)\frac{\gamma_{0+}}{2},
\eea
\bea
\lefteqn{H'_{\rm int}=-\frac{1}{2 L_{\rm q}}\eh^2}\nn
&&\times
\Bigl\{
{\left(L_{\rm q}\alpha E_{\rm J}\sin[\gamma_{\rm q-}
-2 \pi f_{\rm q}]+M E_{\rm J0}\cos[\gamma_{0+}/2]\sin[\pi f_{\rm SQ}]\right)}^2\Bigr.\nn
&&\Bigl.-(L_{\rm q}L-M^2){\left(E_{\rm J0}\cos[\gamma_{0+}/2]\sin[\pi f_{\rm SQ}]\right)}^2
\Bigr\}
\label{SHint}
\eea
Here we disregarded $E_{\rm J0}$, $E_{\rm J}$ compared with $(\hbar/(2e))^2/L_{\rm q}$, $(\hbar/(2e))^2/L$
because the inductances ($L$ and $L_{\rm q}$) are small in experimental situations.

\subsubsection{Two-state approximation of the qubit}

To achieve further simplification we have to calculate the quantum states
of the three junction qubit using the Hamiltonian Eq. (\ref{SHq}) with  realistic parameters. 
In the vicinity of $f_{\rm q}=1/2$, Eq. (\ref{SHq}) becomes
\be
H'_{\rm q}=H_{0}+H_{1},
\ee
where
\bea
\lefteqn{
H_{0}=
\he^2\frac{C_{\rm q}}{4}\left(\dot{\gamma}_{\rm q+}^2+\dot{\gamma}_{\rm q-}^2
+2 \alpha \dot{\gamma}_{\rm q-}^2\right)}\nn
&-2 E_{\rm J}\cos[\gamma_{\rm q+}/2]\cos[\gamma_{\rm q-}/2]
+\alpha E_{\rm J}\cos[\gamma_{\rm q-}],&
\eea
and 
\be
H_{1}=\alpha E_{\rm J}\sin[\gamma_{\rm q-}] 2 \pi f,
\ee
with $f\equiv f_{\rm q}-1/2$.  Here, we ignored $f^2$ and higher powers because
$f_{\rm q}$  is always in the vicinity of 1/2 ($0.499 < f_{\rm q}<0.501$) 
in flux-qubit experiments.
Since $H_{0}$ has an even symmetry  about $\gamma_{\rm q-}$ and has potential wells 
on the both sides of $\gamma_{\rm q-}=0$, when $H_{1}=0$, that is, when
$f=0$, the ground state $|g\rangle$
and the first excited state $|e\rangle$ can be expressed as
\bea
|g\rangle_0=(|{\rm L}\rangle+|{\rm R}\rangle)/\sqrt{2},\nn
|e\rangle_0=(|{\rm L}\rangle-|{\rm R}\rangle)/\sqrt{2},
\label{LRB}
\eea
where $|{\rm L}\rangle$ and $|{\rm R}\rangle$ are wavefunctions approximately
localized in the left  and  right wells, respectively.
When other excited states are adequately higher than these states in terms of energy,
this is a two-state system.

For finite but very small $f$, we can carry out a second-order perturbation calculation 
by adopting $H_{1}$ as a perturbation
because $f=f_{\rm q}-1/2$ is very small.
Then we obtain the perturbed ground and excited states by the first order of perturbations as
\bea
\lefteqn{|g'\rangle=|g\rangle_0}\nn
&-2\pi f \alpha E_{\rm J}\left(
\frac{_0\langle e|\sin[\gamma_{\rm q-}]|g\rangle_0}{\Delta}|e\rangle_0
+\sum_i\frac{_0\langle i|\sin[\gamma_{\rm q-}]|g\rangle_0}{E_{i}}|i\rangle_0\right), &\nn
\lefteqn{|e'\rangle=|e\rangle_0}\nn
&-2\pi f\alpha E_{\rm J}\left(
-\frac{_0\langle g|\sin[\gamma_{\rm q-}]|e\rangle_0}{\Delta}|g\rangle_0
+\sum_i\frac{_0\langle i|\sin[\gamma_{\rm q-}]|g\rangle_0}{E_{i}-2 \Delta}|i\rangle_0\right), &\nn
\label{pte}
\eea
where $|i\rangle$ and $E_{i}$ is a second or higher excited state and its energy when there is no 
perturbation.
The qubit is always operated under the condition that the second and higher excited states 
are energetically far from $|g\rangle_0$ and $|e\rangle_0$, $|E_i|,|E_i-2\Delta|\gg \Delta$ are always 
satisfied. Therefore, the third terms on the right hand side of Eq. (\ref{pte}) and the second and
higher order contributions can
be ignored. Using Eq. (\ref{LRB}), we obtain
\be
\left(
\begin{array}{c}
|g'\rangle\\
|e'\rangle
\end{array}
\right)=\frac1{\sqrt{2}}
\left(
\begin{array}{cc}
1+d & 1-d \\
1-d & 1+d
\end{array}
\right)
\left(
\begin{array}{c}
|{\rm L}\rangle\\
|{\rm R}\rangle
\end{array}
\right),
\label{fLRB}
\ee
where
\bea
\lefteqn{d= \frac{2\pi(f_{\rm q}-1/2)\alpha E_{\rm J}}{\Delta}\langle{\rm L}|\sin[\gamma_{q-}]|{\rm L}\rangle}\nn
\lefteqn{
\displaystyle{=\frac{2\pi(f_{\rm q}-1/2)\alpha E_{\rm J}}{\Delta}}}\nn
&\times \int\int
\psi^*_{\rm L}(\gamma_{q+},\gamma_{q-})\psi_{\rm L}(\gamma_{q+},\gamma_{q-})\sin[\gamma_{q-}]
\id\gamma_{q+}\id\gamma_{q-}& \nn
\lefteqn{\equiv \varepsilon(f_{\rm q})/2}&
\label{ddd}
\eea
and
$\psi_{\rm L}(\gamma_{q+},\gamma_{q-})\equiv\langle\gamma_{q+},\gamma_{q-}|{\rm L}\rangle$.
Equation (\ref{fLRB}) clearly shows that the ground and the first excited  states can be expressed
as superposition of $|{\rm L}\rangle$and $|{\rm R}\rangle$ even when the external flux
deviates slightly from the degenerate point $f_{\rm q}=1/2$.

Therefore, we can simplify the qubit Hamiltonian $H'_{\rm q}$ as
\be
H'_{\rm q}=\varepsilon(f_{\rm q}) \sigma_z -\Delta \sigma_x
\label{qsh}
\ee
where 
$\sigma_z$ and $\sigma_x$ are the spin operators with the basis $\{|{\rm L}\rangle, |{\rm R}\rangle\}$.
Although the above relations, Eqs. (\ref{ddd}) and (\ref{qsh}), were obtained perturbatively,
their validity is confirmed by numerical calculations when $\varepsilon$ is comparable to $\Delta$.

This simplification is not solely for calculation convenience;  it also has a significant
meaning in relation to
the measurement process. Although in the original interaction Hamiltonian of Eq. (\ref{SHint})
it appears that the SQUID measures the phase $\gamma_{\rm q-}$, on the real energy scale,
only the information about the qubit, $|{\rm L} \rangle$ or $|{\rm R} \rangle$,
is transmitted to the SQUID.
Therefore, it is now clear that 
the distribution of the switching currents obtained by enormous numbers of measurements
never reflects the probability distribution of the qubit wavefunction 
$|\psi(\gamma_{\rm q-})|^2\equiv |\langle \gamma_{\rm q-}|\psi_{\rm q-}\rangle|^2$.

For further calculations, the interaction Hamiltonian Eq. (\ref{SHint}) should be expressed 
by this two-state approximation for the qubit. The qubit variable appearing in  Eq. (\ref{SHint})
is $\sin[\gamma_{\rm q-}]$ and this should be expressed by the spin operators.
Now,  suppose that
\bea
\langle{\rm R}|\sin[\gamma_{\rm q-}]|{\rm R}\rangle&=&
-\langle{\rm L}|\sin[\gamma_{\rm q-}]|{\rm L}\rangle\equiv \gamma_{\rm b},\nn
\langle{\rm L}|\sin[\gamma_{\rm q-}]|{\rm R}\rangle
&=&\langle{\rm R}|\sin[\gamma_{\rm q-}]|{\rm L}\rangle=0,
\eea
are given. This shows that we can rewrite 
$\sin[\gamma_{\rm q-}]\rightarrow \gamma_{\rm b}\sigma_z$.
$\gamma_{\rm b}$ is a constant of the order of unity, which is 
determined by the $\alpha$ and $f_{\rm q}$ values of the qubit parameters.




Finally, we obtain a simplified Hamiltonian for the qubit-SQUID coupled system
\bea
\lefteqn{
H=\varepsilon(f_{\rm q})\sigma_z -\Delta \sigma_x} \nn
& &+\frac{C_{\rm 0}}{4}\he^2{\dot{\gamma}_{\rm 0+}}^2
-2 E_{\rm J0}\cos[\gamma_{\rm 0+}/2]\cos[\pi f_{\rm SQ}]\nn
& &-\he I_{\rm b}(t)\frac{\gamma_{0+}}{2}\nn
& &
-\frac{1}{2 L_{\rm q}}\eh^2\Bigl\{\left(
L_{\rm q}\alpha E_{\rm J}\gamma_{\rm b}\sigma_z
+M E_{\rm J0}\cos[\gamma_{0+}/2]\sin[\pi f_{\rm SQ}]
\right)^2\Bigr.\nn
 & &\Bigl.-(L_{\rm q}L-M^2){\left(E_{\rm J0}\cos[\gamma_{0+}/2]\sin[\pi f_{\rm SQ}]\right)}^2 \Bigr\}.
\nn
\label{LHam}
\eea

\section{Time evolution of the density operator $\rho(t)$}

\subsection{Density operator for the qubit and dc-SQUID coupled system}

The simplified Hamiltonian for the qubit-SQUID coupled system contains two quantum variables,
the qubit variable $\sigma$  and the SQUID variable $\gamma_{0+}$.
Therefore, the density operator of the coupled system is defined with the basis of
$|\gamma_{0+},\sigma\rangle$ and
is expressed as\cite{nakano}
\begin{equation}
\rho(t)=\int {\rm d}\gamma_{0+}\int {\rm d}\gamma'_{0+}\sum_{\sigma,\sigma'={\rm L},{\rm R}}
\rho_{\sigma \sigma' \gamma_{0+},\gamma'_{0+}}(t)
|\gamma_{0+},\sigma \rangle \langle \gamma'_{0+},\sigma'|.
\label{dop}
\end{equation}
We use Feynman-Vernon's real-time path-integral method \cite{FV}.
We calculated 
the time evolution of the density operator of the $\gamma_{0+}$-$\sigma$ composite system
Using the unitary evolution including the interaction in Eq. (\ref{LHam}). 
Since the bias current $I_{\rm b}(t)$ changes with time, even the unitary evolution
is calculated numerically with the forward Euler method.
Although the $\gamma_{0+}$ value of the SQUID is a continuous variable, it is expressed by
a discrete variable representation with a $31 \sim 41$ basis. The 
prepared bases are the $\gamma_{0+}$ eigenstates when  restricting the kinetic energy and the range of 
$\gamma_{0+}$.

Every pure state of the composite system can be expressed as
\begin{equation}
|\psi_i\rangle = \int\id\gamma_{0+} 
\left\{
 \phi_{{\rm L}i}(\gamma_{0+})|\gamma_{0+}\rangle\otimes |{\rm L}\rangle
+\phi_{{\rm R}i}(\gamma_{0+})|\gamma_{0+}\rangle\otimes |{\rm R}\rangle 
\right\},
\label{pure}
\end{equation}
where $\left| {\gamma_{0+} } \right\rangle $ is the $\gamma_{0+} $ eigenstate, 
and $\phi _{L} (\gamma_{0+} )$ and $\phi _{R} (\gamma_{0+} )$ are 
the coefficients including the amplitude (i.e., not normalized). 
A mixed state is expressed as 
\begin{equation}
\rho(t)=\sum_i p_i|\psi_i\rangle\langle\psi_i|,
\label{rhodef}
\end{equation}
where $p_i$ is the eigenvalues for the eigenstate $|\psi_i\rangle$ of the mixed state density operator.

In order to show the switching current distribution, we plot the time evolution
of the distribution function of 
\bea
p(\gamma_{0+})\equiv 
\langle \gamma_{0+}|{\rm Tr}_{\sigma}[\rho(t)]|\gamma_{0+} \rangle, \nn
=p_{\rm L}(\gamma_{0+})+p_{\rm R}(\gamma_{0+}),
\eea
where
\bea
p_{\rm L}(\gamma_{0+})&=&\sum_i p_i |\phi_{{\rm L}i}(\gamma_{0+})|^2 \nn
p_{\rm R}(\gamma_{0+})&=&\sum_i p_i |\phi_{{\rm R}i}(\gamma_{0+})|^2, \nn
\phi_{{\rm L}i}(\gamma_{0+})&=&\left(\langle\gamma_{0+}|\otimes\langle{\rm L}|\right)|\psi_i\rangle,\nn
\phi_{{\rm R}i}(\gamma_{0+})&=&\left(\langle\gamma_{0+}|\otimes\langle{\rm R}|\right)|\psi_i\rangle.
\label{PLPR}
\eea
and ${\rm Tr}_{\sigma_z}[\cdots]$ means 
taking the trace over the qubit states to obtain a reduced density operator for
$\gamma_{0+}$.  
The switching current distribution $P_{\rm sw}(I_{\rm b})$, which corresponds to the experimentally
obtained distribution can be calculated from the 
portion of the SQUID wavefunction remaining in the potential well
$p_{\rm in}(I_{\rm b}(t))=
\int_{\rm in}{\rm d}\gamma_{0+}\langle \gamma_{0+}|{\rm Tr}_{\sigma}[\rho(t)]|\gamma_{0+} \rangle
$
 by utilizing the following relation
\begin{equation}
P_{\rm sw}(I_{\rm b})=-A\frac{{\rm d} p_{\rm in}}{{\rm d} I_{\rm b}}=
-A \frac{\id p_{\rm in}(t)}{\id t}/\frac{\id I_{\rm b}(t)}{\id t},
\label{PIS}
\end{equation}
where $A$ is a normalization constant.
Here, the region ``in the well'' is defined as $-1.5 \pi <\gamma_{0+}/2<\pi$. Therefore,
$\int_{\rm in}\id \gamma_{0+}$ actually means the integration over the region. 
The upper limit $\gamma_{0+}/2=\pi$ was chosen because the 
energy barrier for the escape of the phase $\gamma_{0+}$ from
the Josephson potential  always has its peak value in the  vicinity of $\gamma_{0+}/2=\pi$
regardless of the amplitude of bias current. The speed $\id I_{\rm b}(t)/\id t$ of the increase in 
the bias current is set at a constant value in  examples discussed in the next section.

\subsection{Effect of decoherence}

Since we calculate  the time evolution of the density operator of the qubit-SQUID coupled system,
the effect of some types of decoherence can be easily taken into account.
Here, we use a simple model of the decoherence, so called, the system-Boson model\cite{C-L,Weiss}.
The Hamiltonian for the system and the Boson environment is given by
\be
H_{\rm tot}=H+H_{\rm B}+g A_{\rm S}\sum_{k}\sum_k (b_k^\dag+ b_k)+\sum_k\omega_k b_{k}^\dag b_{k},
\ee
with the interaction
Hamiltonian $H_{\rm SB}=\alpha \sigma_z\sum_k (b_k^\dag+ b_k)$, 
where $H$ is the Hamiltonian of the qubit-SQUID composite system
given in Eq. (\ref{LHam}), $b_k$ is the annihilation operator
of the Boson for the mode $k$, and $g$ is the coupling constant,
and $A_{\rm S}$ is the operator of a physical observable in the qubit-SQUID coupled system,
which interacts with the environment. 
The time evolution 
of the density operator expressed by the influence functional manner, and it can be
calculated numerically by, for example, the program established by Makri's group\cite{Makri}.
However, when  it is allowed to approximate the effect of the interaction with the environment
as a Markovian process, a simpler calculation can be performed, with the Lindblad equation\cite{Lindblad}.
Although if the effect of the non-Markovian character of the system-environment interaction is
important, the Lindblad calculation might not be very useful\cite{Tian,Wilhelm}, however, 
it gives suggestive results, as shown later,
when the details of the origins of the decoherence in the experiments are unknown, 
as in the present case.

The Lindblad equation is given by
\bea
\lefteqn{\frac{\rm d}{{\rm d}t}\rho(t)=\frac{1}{i \hbar}\left[H,
\rho(t)\right]} \nn
&+ \frac{\Gamma}{2}\left(2 A_{\rm S} \rho(t) A_{\rm S} 
- \rho(t) A_{\rm S} A_{\rm S} -A_{\rm S} A_{\rm S} \rho(t)\right),&
\label{ld1}
\eea
where $\Gamma$ is the coupling strength which is proportional to the square of $g$.

When (i) $A_{\rm S}\propto \sigma_z$, the decoherence destroys the quantum superposition
between the $|L\rangle$ and $|R\rangle$ qubit states.
This situation corresponds to the decoherence caused by the 
fluctuation of the external magnetic flux $\Phi_{\rm q}=f_{\rm q}\Phi_0$.
(ii) $A_{\rm S}\propto \gamma_{0+}$ corresponds to the influence of
the bias current $I_{\rm b}$ the dissipation
at normal resistance outside the SQUID. 
Moreover, (iii)
$
A_{\rm S}\propto \partial \gamma_{0+}/\partial t,
$ 
expresses the detection
of the finite voltage that appears across the SQUID.
We will show the influence of this decoherence on the qubit-SQUID coupled system and
the resulting switching currents by  numerical calculations.

\section{Results and discussion}

\subsection{Time evolution of the qubit-SQUID composite system during the measurement}

We carried out numerical calculations of the time evolution of the density
operator given by Eq. (\ref{dop}) of the qubit-SQUID coupled
system during the measurement. 

\begin{figure}
\begin{center}
\includegraphics[width=5.5cm]{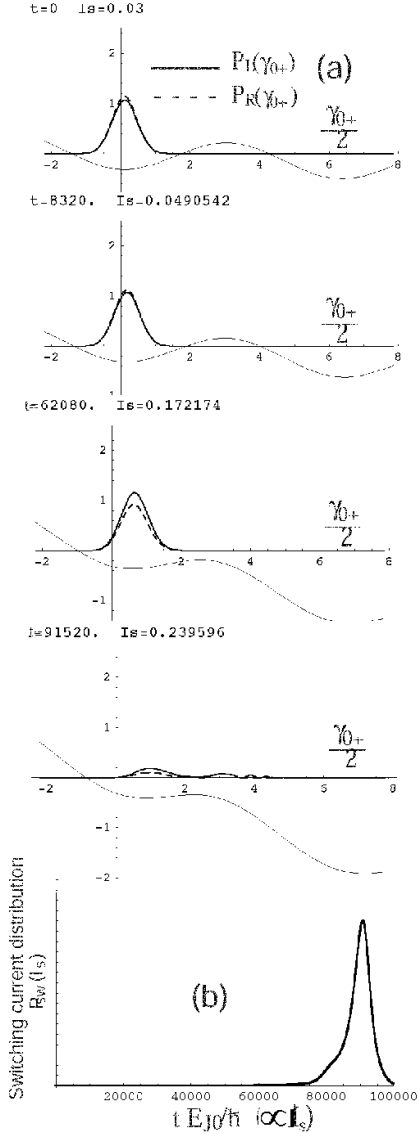}
\end{center}
\caption{
Time evolution of the SQUID probability distribution without decoherence.
$\Delta=0.01 E_{\rm J}$.
(a) Changes in the shape of the probability distribution. 
The solid curves show $p_{\rm L}(\gamma_{0+})$ and the dashed ones show
$p_{\rm R}(\gamma_{0+})$ defined by Eq. (\ref{PLPR}). (b) Switching current distribution calculated by Eq. (\ref{PIS}). 
The time $t$ is proportional to the bias current
$I_{\rm b}(t)$, and normalized by the SQUID Josephson energy $\hbar/E_{\rm J0}$.
$\id I_{\rm b}/\id t= 400$ nA/($\mu$ sec).}
\label{tev-21}
\end{figure}

First, in Fig. \ref{tev-21} we show the time evolution without decoherence, in other words,
the pure unitary evolution of the $\gamma_{0+}$-$\sigma_z$ composite system.
Time $t$ is normalized by the inverse of the Josephson energy $E_{\rm J0}$.
The parameters for the numerical calculations are as follows.
The Josephson energy and the capacitance of a SQUID junction are $E_{\rm J0}=100 $GHz, and
$C_{0}=40$ fF, respectively. This gives $E_{\rm C}/E_{\rm J0}\sim 0.01$.
The inductance of the SQUID ring $L = 10$ pH. The magnetic flux piercing the SQUID
ring is $\Phi_{\rm SQ}=f_{\rm SQ}\Phi_0$ with $f_{\rm SQ}=0.4$.
For the qubit, values of $\varepsilon=-0.0065 E_{\rm J0}$, and $\Delta=0.01 E_{\rm J0}$ are given. 
These  are almost the same as the experimental parameters used in the experiment by an NTT group\cite{Tanaka}.
We employed the mutual inductance $\gamma_{\rm b}M=1.4 L$, which gives approximately ten times greater
qubit-SQUID coupling than in actual experiments carried out with the geometry
shown in Fig. \ref{NTT}. Such  strong coupling is chosen in order to distinguish
the effect of the qubit state on the SQUID switching current within the precision
of our numerical calculations. This interaction is still
weak and does not change the discussion described later.

This set of parameters corresponds to a situation where $|{\rm L} \rangle$ and $|{\rm R} \rangle$ are
almost degenerate. 
($\varepsilon\neq 0$ because  the qubit-SQUID coupling shifts the degeneracy point.)
At $t=0$, we set the SQUID bias current $I_{\rm b}\sim 0$, and
the $\gamma_{0+}$-$\sigma_z$ composite system is in its ground state, that is, the bonding
state of $|{\rm L}\rangle$ and $|{\rm R}\rangle$.
For $t>0$, $I_{\rm b}$ is  increased much more slowly than
the time scale related to other relevant energies. Actually, in this and later  examples,
the bias current increases at approximately $400$ nA/($\mu$ sec). This corresponds to
the bias current reaching the switching current from zero in a time of about 0.1 $\mu$ sec. 

The thin solid curves in Fig. \ref{tev-21} show the Josephson potential for $\gamma_{0+}$ at each time.
The thick solid curves show $p_{\rm L}(\gamma_{0+})$, and
the thick dashed curves show $p_{\rm R}(\gamma_{0+})$, which are defined in Eq. (\ref{PLPR}).
The initial state has amplitudes comparable for both the solid and  dashed curves.
This means that the initial qubit state is a superposition of $|{\rm L}\rangle and |{\rm R}\rangle$.
At this time, since $\phi_{{\rm L}}(\gamma_{0+}) \propto \phi_{{\rm R}}(\gamma_{0+})$ is
approximately satisfied, the qubit and the SQUID are not entangled.
This is always the case when $I_{\rm b}=0$ regardless of the values of the parameters chosen.

\begin{figure}
\begin{center}
\includegraphics[width=5.5cm]{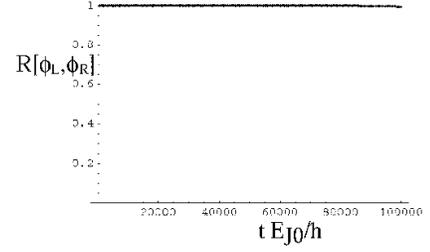}
\end{center}
\caption{
Time evolution of the measure $R\left[\phi_{\rm L},\phi_{\rm R}\right]$ 
of the entanglement between the qubit and the SQUID.
$R=1$ means no entanglement.}
\label{ERfunc}
\end{figure}

With increasing $I_{\rm b}$,
the difference  between 
$|{\rm L}\rangle$ and $|{\rm R}\rangle$ affects the position of the potential minimum in 
the coordinate $\gamma_{0+}$.
As shown in the figure, without decoherence 
the relation $\phi_{{\rm L}}(\gamma_{0+}) \propto \phi_{{\rm R}}(\gamma_{0+})$
is still largely maintained.
Figure \ref{ERfunc} shows this fact quantitatively.
The 
quantity 
\be
R\left[\phi_{\rm L},\phi_{\rm R}\right] \equiv 
\frac{{\left|\int\phi_{\rm L}^*(\gamma_{0+})\phi_{\rm R}(\gamma_{0+})\id\gamma_{0+}\right|}^2
}
{
\int\id\gamma_{0+}{|\phi_{\rm L}(\gamma_{0+})|}^2
\int\id\gamma_{0+}{|\phi_{\rm R}(\gamma_{0+})|}^2
}
\ee 
for a pure state 
measures the degree of entanglement between the qubit and the SQUID. 
$R[\psi _{L} ,\psi _{R} ] = 1$, and 0 indicate no entanglement (separable) and 
maximum entanglement, respectively.

We can see that the quantity remains almost unity as the
SQUID bias current $I_{\rm b}$ increases.  By using this quantity, the so called ``entanglement of formation''
is expressed as
\bea
\lefteqn{E\equiv S(\rho_{\rm SQ})=S(\rho_{\rm qubit})=}\nn
& &2
\int\id\gamma_{0+}{|\phi_{\rm L}(\gamma_{0+})|}^2
\int\id\gamma_{0+}{|\phi_{\rm R}(\gamma_{0+})|}^2
\left(1-R\left[\phi_{\rm L},\phi_{\rm R}\right]\right),\nn
\eea
where $\rho_{\rm SQ}={\rm Tr}_{\sigma}[\rho]$ and $\rho_{\rm qubit}={\rm Tr}_{\rm SQ}[\rho]$, and $S(\rho')\equiv {\rm Tr}[\rho' \log \rho']$
is the von Neuman entropy.
Since $2
\int\id\gamma_{0+}{|\phi_{\rm L}(\gamma_{0+})|}^2
\int\id\gamma_{0+}{|\phi_{\rm R}(\gamma_{0+})|}^2 < 1/2$ in Figs. \ref{tev-21}, and \ref{ERfunc},
the entanglement of formation $E$ is always kept below 0.1\%.

The correlation between $\sigma_z$ and $\gamma_{0+}$ is formed by the interaction
energy between the qubit and $\gamma_{0+}$. However,
the entanglement makes it difficult to utilize the transfer energy $\Delta$
between $|{\rm L}\rangle$ and $|{\rm R}\rangle$ of the qubit because 
the SQUID wavefunctions $\phi_{{\rm L}}(\gamma_{0+})$ and $\phi_{{\rm R}}(\gamma_{0+})$ are different
when entangled, and the overlap between them becomes small. 
It is postulated that this fact suppresses the entanglement
formation. In this situation the switching current may have an intermediate value
between the values for $|{\rm L}\rangle$ and $|{\rm R}\rangle$.
The switching current distribution is shown in Fig. \ref{tev-21}(b).
It has only one significant peak.

\subsection{What do we measure with a dc-SQUID ?}
\label{WDN}

In the vicinity of a half integer of $f_{\rm q}$, the ground state $|g\rangle$ and the first excited state
$|e\rangle$ of the qubit are 
\be
|g\rangle=a(f_{\rm q})|{\rm L}\rangle+b(f_{\rm q})|{\rm R}\rangle,
\hspace{2mm}
|e\rangle=b(f_{\rm q})|{\rm L}\rangle-a(f_{\rm q})|{\rm R}\rangle,
\ee
respectively. Here, we define
\bea
a(f_{\rm q})=\frac{\sqrt{{\varepsilon(f_{\rm q})}^2+\Delta^2}+\varepsilon(f_{\rm q})}
{\sqrt{D}}, \hspace{2mm}
b(f_{\rm q})=\frac{\Delta}{\sqrt{D}}, \nn
D=\Delta^2+
{\left(\varepsilon(f_{\rm q})+\sqrt{\varepsilon(f_{\rm q})^2+ \Delta^2}\right)}^2.
\eea
Since $\varepsilon(f_{\rm q})=0$ at  $f_{\rm q}=1/2$,
$|g\rangle$ and $|e\rangle$ are superpositions of $|{\rm L}\rangle$ and $|{\rm R}\rangle$ at 
$f_{\rm q} \sim 1/2$.
Therefore, $\sigma_z$, which is proportional to the flux induced by the qubit circulating current,  has 
large quantum fluctuations, and does not have a definite value.
Then, what the dc-SQUID switching current $I_{\rm sw}$ measuring the induced flux corresponds to 
is ambiguous. 

A simple explanation of the $\sigma_z$ projection measurement is as follows. 
It requires not only classical correlation but also quantum entanglement 
between the SQUID and the qubit.
When the interaction between the qubit and the SQUID 
produces entanglement between them, as given
by Eq. (\ref{pure}), the wavefunctions 
$\phi_{{\rm L}i}(\gamma_{0+})$ and $\phi_{{\rm R}i}(\gamma_{0+})$ have
their peaks at different positions, $\gamma_{\rm 0L}$ and $\gamma_{\rm 0R}$.
As the SQUID bias current is increased, and
when one of their wavefunctions reaches its switching point, it
switches probabilistically. Therefore, if the two peaks are adequately separated in 
the coordinate $\gamma_{\rm 0}$, we may find two peaks in the switching current distribution.

\begin{figure}
\begin{center}
\includegraphics[width=5.5cm]{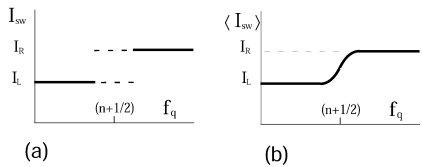}
\end{center}
\caption{
Schematic of dc-SQUID switching current distribution according to a ``projection" measurement
of the qubit.
(a)Before  averaging. (b)Average value.}
\label{heikin}
\end{figure}

We denote the $I_{\rm sw}$ for $|{\rm L}\rangle$ and  $|{\rm R}\rangle$
as $I_{\rm L}$ and $I_{\rm R}$, respectively.
According to a simple measurement formulation, the above  is a $\sigma_z$
measurement. Then, if the simple formulation were correct, the measurement should give
\begin{itemize}
\item
$I_{\rm L}$ with the probability $|a(f_{\rm q})|^2$, 
$I_{\rm R}$ with the probability $|b(f_{\rm q})|^2$.
\end{itemize}
This is a probabilistic phenomenon, and therefore, the obtained values have a large distribution.
These two peaks positioned at $I_{\rm L}$ and $I_{\rm R}$ appear
in the plot of a huge number of obtained values.
After averaging the values, the expectation value
\be
\langle I_{\rm sw}\rangle=\langle g|\hat{I}_{\rm sw}|g\rangle
=|a(f_{\rm q})|^2I_{\rm L}+|b(f_{\rm q})|^2I_{\rm R}
\ee 
would be obtained.
This is shown schematically in Fig. \ref{heikin}.

If the measurement with the SQUID were a  projection measurement of $\sigma _z $ as 
described above, the observed switching current would appear at a value 
corresponding to $\phi_{\rm L} (\gamma_{0+} )$ or 
$\phi_{\rm R} (\gamma_{0+})$, probabilistically. Therefore, if $\phi_{\rm L} (\gamma_{0+} )$ were 
not proportional to $\phi_{\rm R} (\gamma_{0+} )$ and $R\left[{\phi_{\rm L} ,\phi_{\rm R}}\right] 
\ll 1$, the switching current would 
split into two values even for a one qubit state at absolute zero 
temperature, as sketched in Fig. \ref{heikin}(a).

In the experiment reported by the Delft group\cite{Casper},  
(perhaps for a reason that has nothing to do with qubit quantum fluctuations)
the $I_{\rm sw}$ had a broad distribution, which prevents one even from distinguishing $I_{\rm L}$
and $I_{\rm R}$ on a single measurement basis. Therefore, they showed only the averaged values, which faithfully reproduced
the curve shown in Fig. \ref{heikin}(b).

However, as described above, when we carried out numerical calculations on the time 
evolution of the density operator of the total system, we found that 
the total system wavefunction with low energies does not adopt an entangled state but 
remains almost separable, that is, $R\left[ {\phi _{\rm L} ,\phi_{\rm R} } 
\right] \approx 1$ ($\phi _{L} (\gamma_{0+} ) \propto \phi _{R} 
(\gamma_{0+} ) \equiv \phi (\gamma_{0+} )$). Only the position in the $\gamma 
_{0+}$ coordinate of the wavepacket $\phi(\gamma_{0+} )$ depends on the qubit 
state. The obtained switching current corresponds to that of $\phi 
(\gamma_{0+} )$, which is  single-valued, unlike the 
entangled state mentioned above. In addition, the numerical calculation also 
showed that the switching current has a two-value distribution in the 
presence of strong decoherence, which destroys the qubit superposition, as
we will discussed in Figs. \ref{tev-1}, and \ref{CURVE}. 
Since the qubit circulating current $I_{{\rm cir}} $ is proportional to the 
$z$-component of the qubit spin, it should be written as $I_{{\rm cir}} = 
I_{{c}0} \sigma _z $, under the two-state approximation for the qubit. 
According to quantum mechanics, when the spin is in a superposition state 
$\left| \psi \right\rangle = a\left|{\rm L} \right\rangle + b\left|{\rm R} 
\right\rangle $, each  $\sigma _z $ measurement provides the discrete 
result -1 or 1 probabilistically, and never provides an intermediate value. 
Although the quantum-mechanically averaged value of the circulating current 
for the superposition state $\left| \psi \right\rangle = a\left|{\rm L} 
\right\rangle + b\left|{\rm R} \right\rangle $ is given by $\left\langle {I_{\rm cir} 
} \right\rangle = I_{c0} \left( {2\left| a \right|^2 - 1} \right)$, this 
value is obtained after we measure  $I_{{\rm cir}}$  for the state 
$\left| \psi \right\rangle $ many times and calculate its average value. 

Very recently, the NTT group has succeeded in performing single-shot detection
of the qubit flux by improving the  switching current resolution in  their SQUID\cite{Tanaka}.
The plots of the  raw values of the switching currents clearly show the qubit state transition from
the ground state to the first excited state accompanied by a change in the external magnetic
field $\Phi_{\rm q}=f_{\rm q}\Phi_0$, without any averaging of the data\cite{Tanaka}.
One of the  most interesting facts in their data  is that the switching currents illustrate 
a curve that explicitly shows that the qubit is in a superposition
of $|{\rm L}\rangle$ and $|{\rm R}\rangle$ in the vicinity of  a degenerate point $f_{\rm q}= 1.5$.
In such a region, their switching currents appear at  intermediate values between
the switching currents for $|{\rm L}\rangle$ and  $|{\rm R}\rangle$.
A curve that is  very similar to that in Fig. \ref{heikin}(b) appeared 
from the single shot data.
The behavior of the switching current versus the external flux $f$ in Fig. 
\ref{heikin}(b) is very similar to that of the quantum-mechanically averaged value of 
the qubit circulating current $\langle I_{{\rm cir}} \rangle $ in Fig. 
2(b). At first glance, this is very strange if we adopt the above simple formulation of the 
dc-SQUID measurement, which gives a two peak $I_{\rm sw}$ distribution for a superposed
qubit state.


However, taking an experimentally realistic weak qubit-SQUID interaction into account
in the Hamiltonian of the composite system,
one finds that no entangled state between the qubit and the SQUID is formed.
The time evolution keeps the composite system to be in a direct product state even in the presence of
the interaction;
\be
|\psi_{\rm tot}\rangle=|\psi_{\rm SQ}\rangle\otimes|\psi_{\rm q}\rangle.
\ee
The qubit state affects the shape of the SQUID wavefunction;
\be
\phi(\gamma_{0+})\equiv\langle \gamma_{0+}|\psi_{\rm SQ}\rangle.
\ee
The position of the peak of the wavefunction $\phi(\gamma_{0+})$ varies with the qubit state
although the wavefunction is a wavepacket with a finite width.

\subsection{Variational analysis of the entanglement formation between the qubit and the SQUID}

Entanglement formation in time is too complicated to examine analytically.
Therefore, here, we consider the behavior of the entanglement in the ground state of the qubit-SQUID
coupled system by using a variational calculation.

We adopt the trial wavefunction
\bea
\lefteqn{|\psi_{\rm tri}\rangle= {\left(\frac{2}{\pi a}\right)}^{1/4}\int\Bigl(
b_{\rm L}e^{-\frac{(\gamma_{0+}/2-\gamma_{\rm L})^2}{a}}|\gamma_{0+}\rangle\otimes|{\rm L}\rangle\Bigr.}\nn
& &\Bigl.+b_{\rm R}e^{-\frac{(\gamma_{0+}/2-\gamma_{\rm R})^2}{a}}|\gamma_{0+}\rangle\otimes|{\rm R}\rangle
\Bigr)\id \gamma_{0+},
\eea
for the total system because the SQUID potential is nearly parabolic and the wavefunction in the potential 
should take the ground level. The width of the wave packet is $a$.
We denote the difference between the peak positions of the ${\rm L}$ and ${\rm R}$ components
as $d\equiv (\gamma_{\rm L}-\gamma_{\rm R})/2$, and 
mid point between them by
$\gamma \equiv (\gamma_{\rm L}+\gamma_{\rm R})/2$. From the Hamiltonian (\ref{LHam}), 
the expectation value $\langle \psi_{\rm tri}|H|\psi_{\rm tri}\rangle$ of the energy for the trial 
wavefunction is given by
\bea
\lefteqn{\langle \psi_{\rm tri}|H|\psi_{\rm tri}\rangle
=
\frac{E'_{\rm C}}{a}}\nn
&-&2 e^{-a/8}E_{\rm J0}\cos[\pi f_{\rm SQ}]
\left\{\cos\gamma_0 \cos d \right.\nn
& &\left.-(|b_{\rm L}|^2-|b_{\rm R}|^2) \sin \gamma_0 \sin d\right\}\nn
&+&e^{-a/8}E_{\rm J0}k 
\{
\sin \gamma_0 \sin d-(|b_{\rm L}|^2-|b_{\rm R}|^2)\cos\gamma_0 \cos d
\}\nn
&-&\Delta e^{-2 d^2/a}(b^*_{\rm L}b_{\rm R}+b^*_{\rm R}b_{\rm L})\nn
&-&I_{\rm b}(t)\he\gamma_0-I_{\rm b}(t)\he d(|b_{\rm L}|^2-|b_{\rm R}|^2)\nn
&+&\varepsilon(|b_{\rm L}|^2-|b_{\rm R}|^2),
\label{pHp}
\eea
where
\be
k=\frac{M \alpha}{2 }\eh^2E_{\rm J}\gamma_{\rm b}\sin[\pi f_{\rm SQ}],
\ee
is the dimensionless qubit-SQUID coupling constant, and 
\be
E_{\rm C}'\equiv \frac{(2e)^2}{2C_0}.
\ee
Here $k$ is of the order of $10^{-3}$ when we use the parameters in experiments\cite{Casper,Tanaka}.
Therefore, we disregard the higher powers of $k$ in the calculations below.
Here, we ignored the self-energies
for the qubit and the SQUID induced by the interaction, to simplify  the discussion
of the entanglement formation.

We minimize  Eq. (\ref{pHp}) with respect to the peak positions $\gamma_{\rm L}$, and $\gamma_{\rm R}$, and 
the width of the wavefunction $a$, using an approximation that disregards
$k^2$, $d^3$, $a^3$ and higher powers. Then,
we obtain the minimum 
\bea
\lefteqn{
{\rm Min}_{d, \gamma, a}\langle H \rangle
\approx
\frac{E'_{\rm C}}{a_0}+\varepsilon x -\Delta \sqrt{1-x^2}}\nn
&-&\eh\left((1+x k')\sqrt{{I_0}^2-{I_{\rm b}(t)}^2}+I_{\rm b}(t){\rm arcsin}[I_{\rm b}(t)/I_0]\right),\nn
& &
\eea
where
\be
I_0\equiv \eh e^{-a_0/8}E_{\rm J0}2 \cos[\pi f_{\rm SQ}],
\ee
\be
x\equiv |b_{\rm L}|^2-|b_{\rm R}|^2, \hspace{1mm} \sqrt{1-x^2}=b^*_{\rm L}b_{\rm R}+b^*_{\rm R}b_{\rm L},
\ee
\be
k'\equiv \frac{k}{2\cos[\pi f_{\rm SQ}]}.
\ee
This minimum is realized when the peaks are positioned at
\be
\frac{\gamma_{0+}}{2}=\gamma_{\rm L}=\gamma_0+d_0, 
\hspace{1mm} \frac{\gamma_{0+}}{2}=\gamma_{\rm R}=\gamma_0-d_0,
\ee
where
\be
\gamma_0=\arcsin\left[\frac{I_{\rm b}(t)}{I_0}
\left(1-\frac{k' x \frac{4 \Delta}{a_0}\eh}{\sqrt{1-x^2}\sqrt{{I_0}^2-{I_{\rm b}(t)}^2}
+\frac{4 \Delta}{a_0}\eh}\right)\right],
\label{ppos}
\ee
\be
d_0\approx - k' I_{\rm b}
\frac{
\sqrt{1-x^2}
}
{\left[\sqrt{1-x^2}\sqrt{{I_0}^2-{I_{\rm b}}^2}+\frac{4 \Delta}{a_0}\eh\right]},
\ee
and
\be
a_0\approx\frac{E'_{\rm C}}{\sqrt{(2 E_{\rm J0}\cos[\pi f_{\rm SQ}])^2-{I_{\rm b}}^2\he^2}}.
\ee
This split between peaks $d_0(\Delta \neq 0)$ is much smaller than 
the difference between the peak positions $d_0(\Delta=0)$  for the simple (not superposed) $|{\rm L}\rangle$
and  $|{\rm R}\rangle$. In fact, the ratio is given by
\be
\left|\frac{d_0}{d_0(\Delta=0)}\right|=\frac{\sqrt{1-x^2}\sqrt{{I_0}^2-{I_{\rm b}}^2}
}
{
\sqrt{1-x^2}\sqrt{{I_0}^2-{I_{\rm b}}^2}+\frac{4 \Delta}{a}\eh
}.
\ee
This ratio becomes vanishingly small as the bias current  approaching the switching
current, that is, $I_{\rm b}\rightarrow I_0$. Therefore, when the qubit superposition $\Delta$
exists, the entanglement between the qubit and the SQUID is
negligible, at least just before the switching.
At that time, Eq. (\ref{ppos}) gives the position of the SQUID wavefunction peak 
at 
\be
\gamma_0\approx\arcsin\left[\frac{I_{\rm b}}{I_0}
(1-k' x)\right].
\ee
This coincides with the 
peak position when the qubit has a classically definite spin $\sigma_z=x$.

Finally, we can estimate by an approximate variational calculation
that the peak of the SQUID wavefunction is positioned at 
\be
\sigma_z \rightarrow x=|a|^2-|b|^2=2|a|^2-1,
\ee
when the qubit is in a superposed state 
$|\psi_{\rm q}\rangle=a|{\rm L}\rangle+b|{\rm R}\rangle$ 
unless the state of the qubit undergoes a rapid change.
This coincides with the quantum mechanical average $\langle\psi_{\rm q}|\sigma_z|\psi_{\rm q}\rangle$
of $\gamma_{0+}$. Therefore, in the absence of decoherence, we can interpret this to mean that 
the resulting switching current value, at which the $\gamma_{0+}$ escapes
from the Josephson potential, coincides with the quantum-mechanical average $\langle I_{\rm sw}\rangle$
before  any averaging operation.



\begin{figure}
\begin{center}
\includegraphics[width=5.5cm]{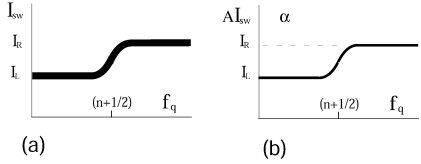}
\end{center}
\caption{
Schematic dc-SQUID switching current distribution according to a "weak" measurement
of the qubit.
(a)Before  averaging. (b)Average value.}
\label{raw1}
\end{figure}

%

\subsection{Effect of decoherence}

\begin{figure}
\begin{center}
\includegraphics[width=5.5cm]{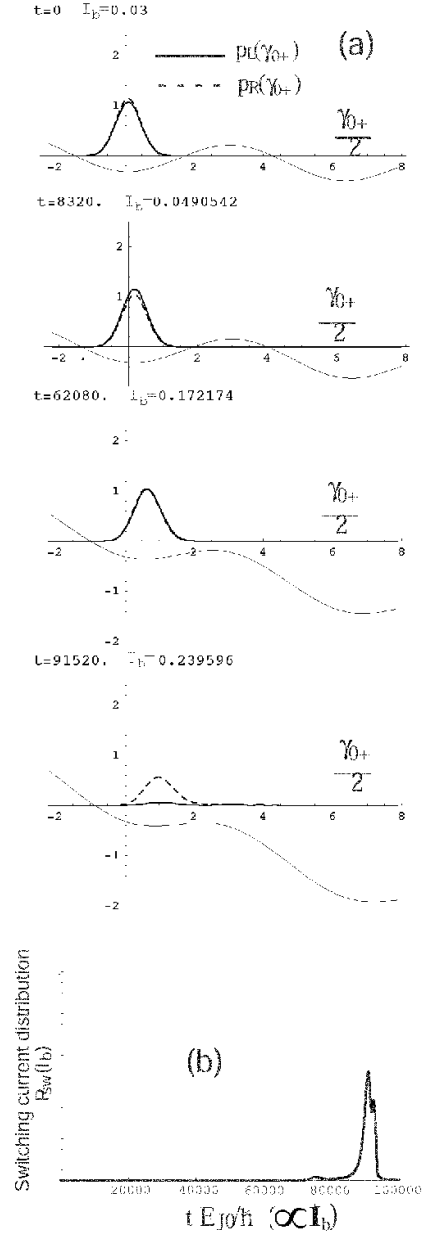}
\end{center}
\caption{
Time evolution of the SQUID probability distribution
with decoherence proportional to $\sigma_z$ with $\Gamma=3.6\times 10^{-4}E_{\rm J0}/\hbar$.
$\Delta=0.001 E_{\rm J}$.
(a) Changes in the shape of probability distribution. The solid curves show $p_{\rm L}(\gamma_{0+})$ and the dashed ones show
$p_{\rm R}(\gamma_{0+})$ defined by Eq. (\ref{PLPR}). 
(b) Switching current distribution calculated using Eq. (\ref{PIS}). 
The time $t$ is proportional to the bias current
$I_{\rm b}$, and is normalized by the SQUID Josephson energy $\hbar/E_{\rm J0}$.
$\id I_{\rm b}/\id t= 400$ nA/($\mu$ sec).
}
\label{tev-1}
\end{figure}

The discussion in Sec. \ref{WDN}
becomes more convincing when we compare the time evolution in Fig. \ref{tev-1}
with  decoherence.

When we use  unrealistically strong coupling between the qubit and the SQUID,
there is a significant entanglement. However, this still gives only one
peak in the switching current distribution in our calculation. This shows that 
entanglement by itself does not cause a two-value measurement, in other words, a $\sigma_z$
projection measurement. Nevertheless, such strongly entangled states
are not robust against a weak decoherence, which destroys the qubit superposition
and realistic switching current distribution for a strong interaction 
as discussed in this section.

Assuming a Markovian process, we performed the calculation using the Lindblad equation (\ref{ld1})
with $A_{\rm S}=\sigma_z$
and $\Gamma=3.6 \times 10^{-4}E_{\rm J0}/\hbar$.
The decoherence makes the composite system a mixed state, and
the states $|{\rm L}\rangle$ and  $|{\rm R}\rangle$ 
become incoherent to each other. 
Therefore, the transfer energy between $|{\rm L}\rangle$ and $|{\rm R}\rangle$
through $\Delta$ does not matter any longer, and the interaction between the qubit and the SQUID 
determines the state of the system 
resulting in the qubit state dependence of the SQUID wavefunction shape.
In this situation, however, the qubit-SQUID correlation is not an entanglement
but a classical correlation
because the states with different qubit states are already incoherent to each other.
In terms of  Eqs. (\ref{rhodef})-(\ref{PLPR}) of the density operator, the ${\rm L}$ component
$\phi_{{\rm L}i}$ and the ${\rm R}$ component $\phi_{{\rm R}j}$ belong to different 
$|\psi_i\rangle$, $|\psi_j\rangle$ ($i \neq j$).
The switching current value becomes one for $|{\rm L}\rangle$ or for $|{\rm R}\rangle$
in a probabilistic manner. This appears in the switching current distribution in Fig. \ref{tev-1}(b).
The peak is split into two, corresponding to $|{\rm L}\rangle$ and $|{\rm R}\rangle$, respectively.

This type of peak splitting in the switching current distribution caused by
the decoherence occurs when the qubit
superposition is weak ($\Delta$ is small), or/and when the qubit-SQUID interaction energy is so
large that a strong qubit-SQUID entanglement is formed before the decoherence makes the system
a mixed state.

Our formulation can readily be used to 
investigate another type of decoherence. Although we do not show the results 
of our numerical calculations here, we  briefly comment on some of them.
When the quantity of the system 
that couples with the environment is the SQUID $\gamma_{0+}$ or $\dot{\gamma}_{0+}$,
the interaction between the system and the environment does not directly deteriorate the qubit
superposition between $|{\rm L}\rangle$ and $|{\rm R}\rangle$. Since the qubit-SQUID coupling through the
mutual inductance is weak in the above calculations (and experiments), the environment
deforms only the shapes of the SQUID wavefunction. This makes the qubit-SQUID system a mixed state.
However, each eigenstate of the mixed-state density operator of the $|{\rm L}\rangle-|{\rm R}\rangle$
superposed qubit is largely maintained even if  the
coupling to the bath is strong. Although the switching current distribution of the mixed-state density 
operator may sometimes  have several peaks, the splitting of the peak does not correspond to 
the splitting of the qubit into $|{\rm L}\rangle$ and $|{\rm R}\rangle$. The peaks instead correspond
to different SQUID states, for example, the ground  and first excited states of the Josephson plasma
oscillation.

\subsection{Switching current behavior with a change in the external magnetic flux $\Phi_{\rm q}$}

\begin{figure}
\begin{center}
\includegraphics[width=5.5cm]{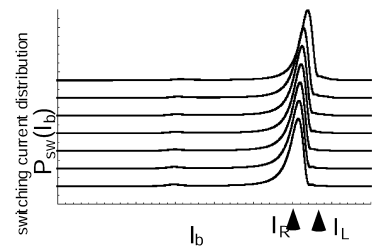}
\end{center}
\caption{
Calculated dc-SQUID switching current distribution in the qubit measurement.
($\Delta=0.01E_{\rm J0}$)
Each curve corresponds to the distribution for an energy difference $\varepsilon$ between 
$|{\rm L}\rangle$ and $|{\rm R}\rangle$. The curves are vertically shifted for clarity. From top to bottom, 
$\varepsilon/E_{\rm J0}=-0.025, -0.015, -0.010, -0.0065, -0.0046, -0.002, +0.005$.}
\label{Dis-P}
\end{figure}

\begin{figure}
\begin{center}
\includegraphics[width=5.5cm]{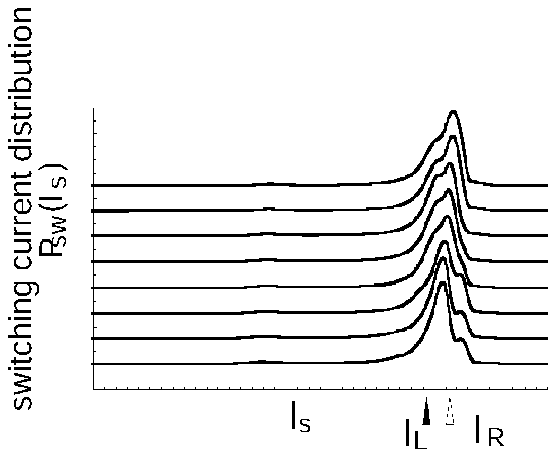}
\end{center}
\caption{
Calculated dc-SQUID switching current distribution in the qubit measurement.
($\Delta=0.01E_{\rm J0}$)
Each curve corresponds to the distribution for an energy difference $\varepsilon$ between 
$|{\rm L}\rangle$ and $|{\rm R}\rangle$. The curves are vertically shifted for clarity. From top to bottom, 
$\varepsilon/E_{\rm J0}=-0.027, -0.025, -0.015, -0.010, -0.0065, -0.004, -0.002,\\+0.005, +0.007$.}
\label{Dis-DP}
\end{figure}

\begin{figure}
\begin{center}
\includegraphics[width=5.5cm]{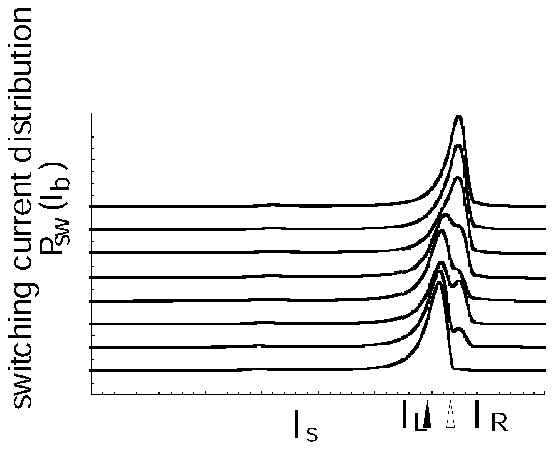}
\end{center}
\caption{
Calculated dc-SQUID switching current distribution in the qubit measurement.
($\Delta=0.001E_{\rm J0}$)
Each curve corresponds to the distribution for an energy difference $\varepsilon$ between 
$|{\rm L}\rangle$ and $|{\rm R}\rangle$. The curves are vertically shifted for clarity. From top to bottom, 
$\varepsilon/E_{\rm J0}=-0.015, -0.010, -0.0065, -0.004, -0.002, 0.000, +0.002,\\
+0.005$.}
\label{Dis-D1}
\end{figure}

If the above discussion of what is indicated about the qubit by the switching current is correct,
our theory must reproduce the switching current behavior 
observed in experiments with the change of the external magnetic flux $\Phi_{\rm q}$
($\propto f_{\rm q}$) shown in Fig. \ref{raw1}. We carried out a time-evolution calculation
at various $f_{\rm q}$ values and present the results in Figs. \ref{Dis-P}, \ref{Dis-DP}, and \ref{Dis-D1}.

Figures \ref{Dis-P} and \ref{Dis-DP} shows the switching current behavior with the energy splitting
$\varepsilon$ between $|{\rm L}\rangle$ and $|{\rm R}\rangle$ when there is no decoherence.
Since $\varepsilon$ is proportional
to the deviation of the external magnetic flux from the degenerate point $f_{\rm q}-(n+1/2)$,
as shown in Eq. (\ref{ddd}), the horizontal axis also corresponds to the change in $f_{\rm q}$.
The peak position of the switching current distribution {\em gradually} shifts from $I_{\rm L}$ 
to $I_{\rm R}$,  and there is no discontinuous transfer. This suggests that the switching current traces
a continuous curve as shown by the schematic illustration in Fig. \ref{raw1}.


\begin{figure}
\begin{center}
\includegraphics[width=5.5cm]{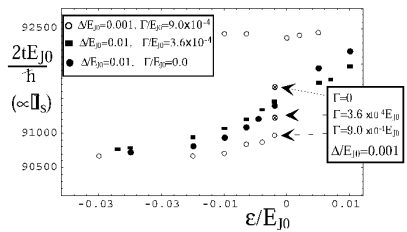}
\end{center}
\caption{$\varepsilon$ dependence of the switching current peak position.}
\label{CURVE}
\end{figure}

Figure \ref{CURVE}, where the peak positions of the switching current distributions
for various external fluxes are plotted,  shows clearly that the 
$\varepsilon$ dependence of $I_{\rm sw}$ obeys such a law. Without decoherence
$I_{\rm sw}$ gradually transfers from the switching current for $|{\rm L}\rangle$ to $|{\rm R}\rangle$ 
across the degenerate point ($\varepsilon=0$). By contrast,
$I_{\rm sw}$ suddenly jumps  from one for $|{\rm L}\rangle$ to $|{\rm R}\rangle$ 
at the degenerate point and two peaks appear at certain flux values.
The  circles indicated by arrows are  the $I_{\rm sw}$ values of the same parameter system for
various decoherence strengths. This provides evidence that
a measurement giving a single intermediate $I_{\rm sw}$ is changed into a two-value measurement
by strong decoherence.

\subsection{Comment on  the  precision of a   single measurement}

The problem analyzed here is invoked by our experiment where  each single switching current measurement
has sufficient resolution to distinguish whether the qubit is in $|{\rm L}\rangle$ or $|{\rm R}\rangle$.
If we could not obtain the switching current behavior depending on the external flux,
it would not make us aware of the importance of understanding the 
switching currents that appear in the intermediate region
between $I_{\rm L}$(for $|{\rm L}\rangle$) and $I_{\rm R}$(for $|{\rm R}\rangle$)
because the averaged value  of the projection measurement results always gives the
intermediate value when the qubit is in a superposition state.
Therefore, here, we briefly consider the factors that determine the resolution of the 
switching currents. 

As we have discussed in this paper, the quantum fluctuations in the qubit-induced flux $\phi_{\rm q}$,
in other words, the quantum fluctuations of $\sigma_z$, 
are not directly related to the SQUID resolution. 
Within the precision of our numerical calculation, there is no significant difference 
between the half width of the switching
current distribution for a $\sigma_z$ definite state and  a superposition state. 
In our numerical calculation, we found
that the distribution width increased when we used a larger SQUID 
junction capacitance. When the capacitance is large, the level splitting of the 
SQUID plasma oscillation of $\gamma_{\rm 0+}$ is narrow, so  the increase in 
the SQUID bias current excites the plasma oscillations. As the result, the state
of $\gamma_{\rm 0+}$ is distributed over many levels that have slightly different 
switching currents. Therefore, the distribution becomes broader resulting in 
a poorer resolution. This fact does not identify the origin of the high resolution 
of our measurement
because there are many other factors that cause those changes in the
distribution width, including the speed of the increase in the bias current, 
and the strength of decoherence. The resolution problem is still an open question.

\section{Conclusion}

We discussed  the superconducting flux-qubit measurement with a dc-SQUID 
from the viewpoint of the dynamics of a composite system consisting of
a qubit and a SQUID.
Since both the qubit and  SQUID are quantum systems the relation
between the measured object (qubit state) and the outcome (switching current) of the measurement apparatus 
is not trivial. 

We calculated the time evolution of the density operator of a qubit-SQUID coupled system from
the onset of an increase in  the bias current to the switching event, and examined
the behavior of the switching current distribution. 

Of particular interest is the fact that when the ground state 
of the composite system is a qubit superposition state,
the measurement designed to distinguish in which state the qubit is does not always work as designed,
and there is a possibility that the switching current
will have an intermediate value between the two states.
Nevertheless, even in such a situation, when there is decoherence
(reduction in the off-diagonal elements of the density operator) which destroys
the qubit superposition of $|{\rm L}\rangle$ and $|{\rm R}\rangle$, 
the switching current then  gives us information about $|{\rm L}\rangle$
or $|{\rm R}\rangle$
in a probabilistic manner.


When the external field $\Phi_{\rm q}$ is far from the degenerate point ($\Phi_{\rm q}=(n+1/2)\Phi_0$),
since there is no quantum transfer between the two states of the qubit,
the qubit should be in the lower energy state, and the switching current of the SQUID
shows a value corresponding to that state.
The switching current $I_{\rm sw}$ is different 
for $\Phi_{\rm q}\ < (n+1/2) \Phi_0$ and $\Phi_{\rm q}\ >  (n+1/2) \Phi_0$
because the energies of the two  states cross 
at the degenerate point.

We find by variational calculation that the peak position of the wavepacket $\phi(\gamma_{0+} )$ is 
approximately the bottom of the SQUID Josephson potential $V_{\rm SQ} (\gamma_{0+} ;\sigma _z = 
2\vert a\vert ^2 - 1)$. The resulting switching 
current becomes that for a usual magnetic field measurement with SQUID of this potential. This value 
coincides with the quantum-mechanical average $\langle I_{\rm sw} \rangle 
$ of the switching current for the qubit $\left| \psi_{\rm q} \right\rangle = 
a\left|{\rm L} \right\rangle + b\left|{\rm R} \right\rangle $. The fact that the 
switching current behaves as the average value, which may at first seem 
strange, is explained as described above. Moreover, this indicates a very 
important fact about the dc-SQUID measurement of flux-qubit states. The 
measurement is not a projection measurement that determines whether the 
qubit is $\left|{\rm L} \right\rangle $ or $\left|{\rm R} \right\rangle $. It is a 
simple measurement of the weight $\vert a\vert ^2( = 1 - \vert b\vert ^2)$ 
of the qubit.

The flux $\phi _{\rm q} $ induced by the qubit ring current has a large quantum 
fluctuation when the qubit is in a superposition state. The fluctuation of 
the measured switching current, however, has almost nothing to do with the 
fluctuation of $\phi _{\rm q} $. The switching current fluctuation is caused mainly by 
fluctuations of the SQUID itself.

Our theoretical results in the present work  clearly explain
the switching current behavior revealed by our experimental results.
This may provide us with important clues to
improve the measurement scheme for quantum computations.

\section*{Acknowledgments}

We thank   S. Ishihara for his encouragement throughout this work.
We also thank  A. Shnirman and 
C. J. P. Harmans, and  D. Esteve, for their helpful comments.
We also acknowledge J. E. Mooij,   C. van der Wal,   Y. Nakamura, and M. Devoret 
for useful discussions.

\appendix

\section{Elimination of variables, $\gamma_{0-}$ and $\gamma_{\rm q3}$}

Here, we briefly explain how we eliminate invisible variables, $\gamma_{0-}$ and $\gamma_{\rm q3}$.

Picking up terms including $\gamma_{0-}$ or $\dot{\gamma}_{0-}$ from the Hamiltonian Eq. (\ref{Tot}),
we obtain
\be
H(\gamma_{0-})=\frac1{2}m {\dot{x}}^2 +\frac1{2}m {\omega_0}^2x^2+xF(t)
-2 E_{\rm J0}\cos[\gamma_{\rm 0+}]\sin[\pi f_{\rm SQ}],
\ee
where
\be
m\equiv\frac{C_0}{2}\he^2,
\ee
\be
x\equiv \gamma_{0-} -2 \pi f_{\rm SQ},
\ee
\be
\omega_0\equiv \sqrt{\frac{2 L_{\rm q}}{C_0(L_{\rm q}L-M^2)}-\eh\frac{2 E_{\rm J0}}{C_0}
\cos[\gamma_{0+}/2]\cos[\pi f_{\rm SQ}]},
\ee
\be
F(t)\equiv \he^2 \frac{M \gamma_{\rm q}(t)}{L_{\rm q}L-M^2}
+E_{\rm J0}\cos\left[\gamma_{0+}/2\right]\sin[\pi f_{\rm SQ}],
\ee
\be
\gamma_{\rm q}\equiv \gamma_{\rm q-}+\gamma_{\rm q3}-2 \pi f_{\rm q}.
\ee
Here, we neglected $x^3$ and higher powers such as $\cos[x]\rightarrow 1-x^2/2$, $\sin[x]\rightarrow x$ because
$x$ is strongly constricted in the parabolic potential. 

Then, the Feynman-Vernon influence functional expressing the effect of $\gamma_{0-}$
on the motion of $F(t)$ is given by\cite{FV,Weiss}
\be
{\cal F}_{\rm FV}=\exp\left[-\Phi_{\rm FV}\right],
\ee
where
\bea
\lefteqn{\Phi_{\rm FV}\equiv
\frac1{\hbar}\int_{0}^{t}\id t_1\int_{0}^{t_1}\id t_2
[F(t_1)-F'(t1)]}\nn
& & \times[L(t_1-t_2)F(t_2)-L^*(t_1-t_2)F'(t_2)],
\label{FV}
\eea
\be
L(t)\equiv
\frac1{2 m \omega_0}
\left(
\coth\left[\frac{\hbar \omega_0 \beta}{2}\right]\cos[\omega_0 t]-i \sin[\omega_0 t]
\right).
\ee

Carrying out the $t_2$ integration in Eq. (\ref{FV}) by part two times, 
\bea
\lefteqn{\int_0^{t_1} \id t_2 \left[L(t_1-t_2)F(t_2)-L^*(t_1-t_2)F'(t_2)\right]} \nn
& &=\frac1{2 m \omega_0^2}
\Biggl[
-i (F(t_1)+F'(t_1)-\cos \omega_0 t_1(F(0)+F'(0))\Biggr.\nn
& &\Biggl.+\coth\left[\frac{\hbar \omega_0 \beta}{2}\right]
\sin \omega_0 t_1 (F(0)-F'(0))
\Biggr] \nn
& & +\frac1{2 m {\omega_0}^3}
\Biggl[
-i \sin \omega_0 t_1 (F(0)+F'(0)) 
+\coth\left[\frac{\hbar \omega_0 \beta}{2}\right]\Biggr. 
\nn
& &\Biggl.
\left(
\cos \omega_0 t_1 (\dot{F}(0)-\dot{F}'(0)) - (\dot{F}(t_1)-\dot{F}'(t_1))
\right)
\Biggr] \nn
& &-\frac1{{\omega_0}^2}
\left[
\int_{0}^{t_1}L(t_1-t_2)\ddot{F}(t_2)\id t_2\right.\nn
& & \left. -\int_{0}^{t_1}L^*(t_1-t_2)\ddot{F}'(t_2)\id t_2
\right],
\label{Line}
\eea
is given. Here, $F(t)$ indicates the quantity on the forward line and $F'(t)$ that on the backward one.
`$\dot{A}$' and `$\ddot{A}$' mean the first and the second time derivatives of the quantity $A$,
respectively.
Since 
$\dot{F}(t)=\he^2 \frac{M }{L_{\rm q}L-M^2}\dot{\gamma}_{\rm q}(t)
-\frac{E_{\rm J0}}{2}\sin\left[\gamma_{0+}/2\right]\sin[\pi f_{\rm SQ}]\dot{\gamma}_{0+}$,
and both $\dot{\gamma}_{\rm q}(t)$ and $\dot{\gamma}_{0+}$ are of the order of
$\sqrt{E_{\rm J}E_{\rm C}}/\hbar$ at most,
the terms from the second line of Eq. (\ref{Line}) are smaller by the factor 
$\sqrt{E_{\rm J}E_{\rm c}}/(\hbar \omega_0)\sim \eh \sqrt{L E_{\rm J0}}$ than 
those in the first line.
This  factor is no larger than $10^{-1}$ in experiments\cite{Casper,WalT,Tanaka}, therefore, we only
take the terms in the first line in Eq. (\ref{Line}).

Now we obtain
\bea
\lefteqn{\Phi_{\rm FV} \approx \frac1{2 \hbar m \omega_0^2}\int_{0}^{t}\id t_1
[F(t_1)-F'(t1)]}\nn
& & \times
\Bigl[
-i (F(t_1)+F'(t_1)-\cos \omega_0 t_1(F(0)+F'(0))\Bigr. \nn
& &\Bigl.+\coth\left[\frac{\hbar \omega_0 \beta}{2}\right]\sin \omega_0 t_1 (F(0)-F'(0))
\Bigr].
\eea
Highly oscillatory terms such as, $\sin \omega_0 t_1 $, $\cos \omega_0 t_1$
are suppressed by the integration over $t_1$. This enables us
to neglect these terms.

Finally, we obtain the influence functional 
\bea
{\cal F}_{\rm FV}&\approx&\exp\left[-\frac{i}{2 \hbar m \omega_0^2}\int_{0}^{t}\id t_1
[F(t_1)-F'(t_1)]
[F(t_1)+F'(t_1)]
\right]\nn
&
=&\exp\left[-\frac{i}{2 \hbar m \omega_0^2}\int_{0}^{t}\id t_1
[{F(t_1)}^2-{F'(t_1)}^2]
\right]\nn
& 
=&
\exp\left[-\frac{i}{2 \hbar m \omega_0^2}\int_{0}^{t}
{F(t_1)}^2\id t_1\right]\nn
&\times&
\exp\left[\frac{i}{2 \hbar m \omega_0^2}\int_{0}^{t}
{F'(t_1)}^2\id t_1\right].
\eea
This influence functional is local in time and $F(t_1)$ and $F'(t_1)$ are separated 
so that the effect of $\gamma_{0-}$ gives a unitary evolution which is
expressed by the Hamiltonian
\bea
\lefteqn{H=\frac1{2 m \omega_0^2}F(t)^2}\nn
& & = -\frac{L_{\rm q}L-M^2}{2 L_{\rm q}} \eh^2 \nn
& & \times\left(\he^2
\frac{M \gamma_{\rm q}(t)}{L_{\rm q}L-M^2}
+ E_{\rm J0}\cos\left[\gamma_{0+}/2\right]\sin[\pi f_{\rm SQ}]
\right)^2\nn
& &
\eea

Very similarly, the elimination of $\gamma_{\rm q3}$
is carried out as follows. The Hamiltonian concerning $\gamma_{\rm q3}$ is
\bea
\lefteqn{H(\gamma_{\rm q3})=\frac1{2}m_{y} \dot{y}^2 + \frac1{2}m_y {\omega_1}^2y^2 + F_y(t)y}\nn
& &
-\frac{L_{\rm q}L-M^2}{2 L_{\rm q}}\eh^2 {E_{\rm J0}}^2\cos^2[\gamma_{0+}/2]\sin^2[\pi f_{\rm SQ}]\nn
& &
-\alpha E_{\rm J}\cos[\gamma_{\rm q-}-2 \pi f_{\rm q}] -(\gamma_{\rm q-}-2\pi f_{\rm q})\times \nn
& &\left(\alpha E_{\rm J}\sin[\gamma_{\rm q-}-2\pi f_{\rm q}]
+\frac{M}{L_{\rm q}}E_{\rm J0}\cos[\gamma_{0+}/2]\sin[\pi f_{\rm SQ}]\right)\nn
& &+\left(\frac1{2}\alpha E_{\rm J}\sin[\gamma_{\rm q-}-2\pi f_{\rm q}]-\frac1{2 L_{\rm q}}\he^2\right)
(\gamma_{\rm q-}-2\pi f_{\rm q})^2, \nn
& &
\eea
where
\be
m_y\equiv\alpha C_{\rm q}\he^2,
\ee
\be
y \equiv \gamma_{\rm q3},
\ee
\be
m_y{\omega_1}^2\equiv 
\alpha E_{\rm J}\sin[\gamma_{\rm q-}-2\pi f_{\rm q}]+\frac1{L_{\rm q}}\he^2,
\ee
\bea
\lefteqn{F_y(t)\equiv}\nn
& &(\gamma_{\rm q-}-2\pi f_{\rm q})
\left(
\alpha E_{\rm J}\cos[\gamma_{\rm q-}-2\pi f_{\rm q}]-\he^2\frac1{L_{\rm q}}
\right)\nn
& &
-\frac{M}{L_{\rm q}}E_{\rm J0} \cos[\gamma_{0+}(t)/2]\sin[\pi f_{\rm SQ}]
-\alpha E_{\rm J}\sin[\gamma_{\rm q-}-2\pi f_{\rm q}], \nn
& &
\eea
Here, we neglected $(\gamma_{\rm q3}+\gamma_{\rm q-}-2 \pi f_{\rm q})^3$ and higher powers because
they are strongly constricted in the parabolic potential. 
The influence functional expressing the effect of $\gamma_{\rm q3}$ on $\gamma_{\rm q-}$
and $\gamma_{0+}$ is given from this description. Moreover, the terms that are non-local
in time are negligible for the same reason as that for $\gamma_{0-}$.
Finally, we obtain the simplified Hamiltonian Eq. (\ref{SHq}) where $\gamma_{\rm q3}$ and $\gamma_{\rm 0-}$
have been eliminated.

\section{Discrete Variable Representation of $\gamma_{\rm 0+}$}

Since the variable $\gamma_{0+}$ is a continuous one, it is not directly suitable for
numerical calculations. Therefore, we discretized them as follows.

The domain of $\gamma_{0+}$ is restricted within $(a, b)$, where $a=-0.7\pi, b=2.5\pi $. 
This domain is wide enough
to express the SQUID wavefunction. We prepare orthonormal basis $\{|k_n\rangle\}$, which
satisfies  $\langle\gamma_{0+}|k_n\rangle=\exp[i k_n (\gamma_{0+}-a)]$, where $k_n= 2 n \pi/(b-a)$,
$n=0, \pm 1, \pm 2, \cdots, \pm N$. They are Fourier series and $\dot{\gamma}_{0+}$ eigenstates.
From this set of $\{|k_n\rangle\}$, we make $\gamma_{0+}$ eigenstates 
\be
\hat{\gamma}_{0+}|\gamma_m\rangle=\gamma_m|\gamma_m\rangle, \hspace{2mm}(m=0, \pm 1, \pm 2, \cdots, \pm N),
\ee
where
\be
|\gamma_m\rangle=\sum_n a_{mn}|k_n\rangle, \hspace{2mm}\sum_n {|a_{mn}|}^2=1.
\ee

We use this basis $\{|\gamma_m\rangle\}$ for the discrete variable representation
of $\gamma_{0+}$ and the wavefunction of the SQUID is expressed as the linear
combination of $|\gamma_m\rangle$.
Using this basis with finite $N$ corresponds to the abandonment of 
the states of higher momentum energies in the SQUID. 
Therefore, we can calculate low energy states precisely with small 
numbers of elements of the basis compared with other bases, such as $\gamma_{0+}$ eigenstates placed
with the same interval over  $(a,b)$.

\end{document}